%% file: Ne20Coul5.tex
\documentclass[12pt,a4paper]{article}
\usepackage[]{import}

\import{}{Ne20Russpckge.tex}
\begin{document}
\setlength{\parskip}{0pt}
  \setlength{\parindent}{16pt}
  \addtolength{\skip\footins}{1.5 mm}
\newcommand{\half}{\frac{1}{2}}
  \begin{center}
  {\bf \Large Coulomb-Corrected Wormhole Model for Neon-20 \\ [2.3em]}
   {\bf \large Ruoxin Bai and Nicholas S. Manton\\ [2em]
}
{\emph{Department of Applied Mathematics and Theoretical Physics, \\
    University of Cambridge, Wilberforce Road CB3 0WA, U.K. \\ and \\
  St John's College, Cambridge CB2 1TP, U.K.}}
\end{center}

\vspace{2em plus 0.5ex minus 0.2ex} 

\begin{abstract}
  
Building on the spatial wormhole geometry proposed by Manton and
Dunajski, we develop a modified model for the Neon-20 nucleus that
incorporates a repulsive Coulomb potential. This reduces the large threshold
energy for cluster break-up in the original model and converts most
bound states to resonances. We use generalized WKB methods to
calculate the energies of bound states, and also the energies and widths
of under-barrier and over-barrier resonances. The results align closely with
experimental data for the \(0_{1}^{+}\), \(0^{-}_{1}\) and \(0_{4}^{+}\)
rotational bands of Neon-20, including the large widths
in the higher-nodal \(0_{4}^{+}\) band.
  
\end{abstract}

\vfill

\clearpage
\section{Introduction}

In the alpha-particle model of nuclei having equal, even numbers of
protons and neutrons, Neon-20 is modelled as a triangular bipyramid of
five alpha particles with $D_{3h}$ symmetry. Direct experimental evidence for alpha-particle clustering in the Neon-20 ground state has recently been established \cite{harris}. When the bipyramid is
treated as a quantised rigid-body rotor, the ground-state band is a $0^+$
band, where the angular momentum projection along the $C_3$-axis is
zero. This is identified with the experimental $0^+_1$ band. There are
also higher-energy pure rotational bands with non-zero angular momentum
projections along the $C_3$-axis, e.g. a $3^-$ band, but we ignore these.
The bipyramid is therefore effectively axially symmetric.

Starting in the 1960s, quantised vibrational modes of the bipyramid
have been studied, and recently, Bijker and Iachello have classified
all the low-lying, rotational-vibrational bands having up to two
vibrational phonons \cite{bijker}. Here, we will focus on the non-degenerate
lowest-frequency mode, which tends to split the bipyramid into an
alpha particle and a tetrahedral cluster of four alpha particles, forming
Oxygen-16. This is motivated by the Neon-20 Ikeda diagram \cite{vOFK},
which shows that the asymmetric $4+1$ cluster break-up energy is just
4.73 MeV, whereas the symmetric $3+1+1$ break-up requires nearly
12 MeV. The 0-, 1- and 2-phonon rovibrational bands for this mode can be
identified with the experimental ground-state $0^+_1$ band, the $0^-_1$ band
and the $0^+_4$ band, respectively. The 2-phonon band is
the `higher-nodal' band of Fujiwara et al. \cite{fuji}; in our
model, the vibrational wavefunction has two nodes on opposite sides of
the bipyramid.

In \cite{wormhole} Manton and Dunajski considered arbitrarily large
amplitudes for this lowest mode, and noted that positive
amplitude describes an alpha particle separating from the bipyramid
in one direction, while negative amplitude describes a different alpha
particle separating in the opposite direction (see Fig. 1). An
oscillation of this mode is therefore analogous to the
dynamics of a Newton cradle. As a consequence, the
configuration space, stretching from the bipyramid out to well-separated,
two-cluster configurations with the separation axis in all possible
directions, is geometrically a spatial wormhole.

\begin{figure}[H]
  \centering
  \includegraphics[width=10cm]{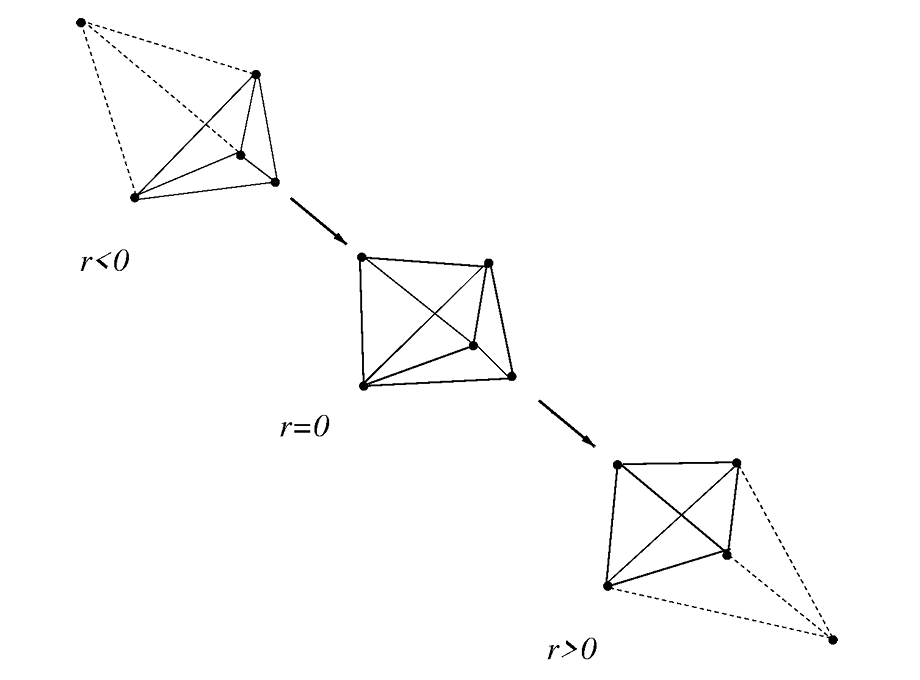}
  \captionsetup{}
  \caption{An incoming alpha particle approaches a tetrahedron of four
  alpha particles, instantaneously forms a $D_{3h}$-symmetric bipyramid, and
  then the opposite alpha particle is ejected.}
\end{figure}

The wormhole is a 3-dimensional manifold acted on by the rotation
group $SO(3)$, where all orbits are 2-spheres. The generic orbits are
those of the separated $4+1$ clusters. The smallest orbit, the
throat of the wormhole, is that of the bipyramid. 
The wormhole has a radial coordinate $r$ and standard spherical polar
coordinates $\theta, \phi$. The radial coordinate runs from $-\infty$ to
$\infty$, because two distinct alpha particles are involved. However,
there is a $\Z_2$ action reversing the sign of $r$ and mapping the angular
coordinates of a point to those of the antipodal point on the 2-sphere;
on the orbit at $r=0$, it simply maps a point to its antipode. The
effect of this $\Z_2$ is to exchange alpha-particle labels, but as
alpha particles are identical bosons, wavefunctions on the wormhole
need to be $\Z_2$-symmetric. The wormhole has a standard
smooth, curved Ellis--Bronnikov metric, given below;
however, far from the throat in both directions, the wormhole flattens
out to be asymptotically Euclidean $\R^3$. This is required by the Euclidean
geometry of two well-separated, effectively pointlike clusters. The
idea that cluster dynamics is nonlinear and leads via inertia
calculations to non-trivial, curved geometry, has been advocated by
Wen and Nakatsukasa \cite{wen}, among others, but the proposed
wormhole geometry in ref.\cite{wormhole} is, we believe, novel.

On the wormhole, a short-range attractive potential is
introduced, modelling the strong nuclear attraction of the clusters.
This potential is spherically symmetric and of harmonic oscillator
type near the bipyramid, with a small anharmonic correction. The
angular part of a wavefunction is classified by an integer
angular momentum $l$, as usual, and the radial part can be classified
by a vibrational phonon number $n$. The bosonic $\Z_2$ symmetry
requires the parities of $n$ and $l$ to be equal.
The 0-phonon, 1-phonon and 2-phonon states in the wormhole model are
consequently rather similar to those in the rovibrational model of
ref.\cite{bijker} and its predecessors.

The short-range potential in \cite{wormhole} is of finite depth,
approaching a constant for infinite cluster separation. The rovibrational
bands are therefore truncated. For each angular momentum $l$, the phonon
number $n$ can be no greater than $8-l$, and there are no
bound states with $l \ge 9$. For example, the ground-state $0^+$ band,
with $n=0$, has positive-parity bound states with $l=0,2,4,6,8$. The
next higher $0^-$ band, with $n=1$, has negative-parity states
with $l=1,3,5,7$. The energies of these states (above the ground
state) reach up to a break-up threshold of just over 15 MeV.

The main problem with this original wormhole model of Neon-20, and of some of
the earlier rovibrational models, is that it takes no account of the
long-range Coulomb repulsion between clusters. This dominates for large
separations and lowers the experimental break-up threshold to 4.73 MeV.
In this paper, we retain the wormhole geometry and the short-range
potential, but add a repulsive Coulomb potential. We slightly change
the coefficients from those in \cite{wormhole} in order to maintain a
good fit to data for the real parts of the energies in
the ground-state band. The combined potential has a peak at a
certain positive radius, and as it is symmetric in $r$, it is actually
double-humped. A few states in the 0-phonon band have
real energies below the new break-up threshold, so remain as bound
states, but the others are above this threshold and become finite-width
resonances. States with real energies below the effective
potential barrier height (which includes an $l$-dependent centrifugal
contribution) have relatively narrow widths. States just above the
barrier have large widths. We will calculate all these energies and
widths using WKB methods. The main result of this paper is Table 4
below, presenting the real energies and widths for states in the
0-, 1- and 2-phonon rovibrational bands. We find that there is a
satisfactory, though not perfect, match to experimental data.

In section 2 we set up the Coulomb-corrected wormhole model. In sections 3 and
4 we review the WKB methods we use, and check them in the context of the
original wormhole model without the Coulomb potential. In section 5
we determine the optimal parameters for the Coulomb-corrected model,
and in section 6 calculate its spectrum of energies and widths, and compare
with experimental Neon-20 data. Section 7 presents our conclusions.

\vspace{1em plus 0.5ex minus 0.2ex} 

\section{Wormhole Model Including a
  Coulomb Potential}

In conventional models, a Euclidean configuration space $\R^3$ is used for
the separation vector between two clusters, often with a central
exclusion zone or very strong repulsive potential to prevent cluster
overlap. However, this leads to singularities. In contrast, Manton
and Dunajski's Neon-20 model has a smooth, curved wormhole as its
3-dimensional configuration space \cite{wormhole}, whose throat
represents the coalescence of the clusters into a five-alpha-particle
bipyramid. The throat size is determined by the moment of inertia of
the bipyramid about a line orthogonal to its $C_3$-axis.
Additionally, there is a potential energy whose minimum is at the throat.

This original wormhole model is an anharmonic extension of a
rotational-vibrational model for Neon-20, focusing on the singly-degenerate
vibrational mode, transforming under the $A_2''$ representation of the
bipyramid's $D_{3h}$ symmetry, that leads to a $4+1$ alpha-particle
cluster split. The three lowest-lying rovibrational bands -- the 0-phonon,
1-phonon and 2-phonon bands -- are identified with the experimentally
established $0^+_1$, $0^-_1$ and $0^+_4$ bands of Neon-20.
In \cite{wormhole}, the spectrum of the wormhole model was studied using
both an anharmonic oscillator approximation and a numerical method. The
model's energy levels align well with the real energies of the rotational
bands listed above, with a threshold energy of 15.19 MeV
for cluster unbinding.

However, experimental data from the TUNL \cite{tunl} and ENSDF
\cite{ensdf} tables indicate that the break-up energy of Neon-20 is
4.73 MeV. Consequently, only three states from the original model are
true bound states, while those with energies above 4.73 MeV should be
resonances, with complex energies. These states have widths (decay
lifetimes) that cannot be analysed in the original model. To achieve
the correct threshold energy and pattern of widths, we need to extend
the model to incorporate a repulsive Coulomb potential between
clusters, which was previously absent.

We retain the Ellis--Bronnikov wormhole metric
\cite{ellis,bronnikov} on the configuration space,
\begin{equation}\label{metric}
  \d s ^{2}=\d r^{2} + (r^{2}+a^{2})(\d  \theta ^{2}+ \sin ^{2} \theta
  \, \d \phi^{2}) \,.
\end{equation}
The range of $r$ is doubly infinite to allow for an incoming alpha
particle to merge into the bipyramid when $r=0$, and for the opposite
alpha particle to be outgoing, but as alpha particles are bosons there
is the further $\Z_2$ quotient mentioned above.
\( |r| \) asymptotically approaches the physical separation \( R \)
between the centres of the two clusters, i.e., \( R / |r| \sim 1 \)
as \( |r| \longrightarrow  \infty \). The throat radius is $a$.

On the wormhole background, we introduce a modified potential combining
a strong short-range attraction with a long-range Coulomb repulsion,
\begin{equation}\label{Pot}
V_{\rm total}(r) = -\frac{V_0}{(r^{2}+d^{2})^{2}} +
\frac{Q_0}{({r^{\kappa}+b^{\kappa}})^{1/ \kappa}} \,,
\end{equation}
where $V_0$ and $Q_0$ are both positive. We refer to the second term
as the Coulomb potential $V_{\mathrm{Cou}}(r)$. The length parameters
$b$ and $d$ will turn out to be somewhat less than the previously
established wormhole radius $a$. $\kappa$ is positive and dimensionless.
The family of potentials (\refeq{Pot}) all have the same asymptotic Coulomb
behaviour \( Q_0 / r \). The original potential without the Coulomb
term is monotonic, and therefore has only bound states and continuum
states. But the modified potential has two peaks, and therefore has
finite-width resonance states above the break-up threshold. In subsequent
calculations, we will set $\kappa=8$. 

Formally, resonances are identified as poles of the scattering
matrix on the complex energy plane, but calculating the pole positions
can be mathematically tricky and
computationally costly. To simplify, we use generalized WKB approximations.
We will review the standard WKB approximation for bound states
\cite{landau}, and Shepard's improved WKB approximation for under-barrier resonances \cite{shepard}, and check them by
applying them to the original model. Note that the original model's
effective potential has a peak and admits an under-barrier resonance
for $l=10$. We will also extend Shepard's method to
compute energies and widths of over-barrier resonances. Most
recognised states of Neon-20 are bound states or under-barrier
resonances. A few broad, higher-energy states are identified as
over-barrier resonances.

In the original wormhole model, the dimensionless parameter $m$, proportional
to the parameter $V_0$, was fixed at $m = 9$.
In principle, $m$ may assume any positive value; in our new
model we set $m = 8$ to achieve a better fit to experimental
observations. This adjustment effectively lowers the threshold energy
by making the potential well shallower.

Having fixed $m = 8$, the parameters $a$ and $d$ are determined by
a least-squares fit to the real energies of the states from $l=0$ up
to $l=8$ in the ground-state $0^+_1$ band. The Coulomb potential has
a negligible effect on these energies, apart from a constant shift.
Subsequently, $Q_0$ and $b$ are fixed by imposing conditions on
the potential as $r \to \infty$ and $r \to 0$ that ensure
the correct Coulomb tail, and the correct threshold energy 4.73 MeV.

\vspace{1em plus 0.5ex minus 0.2ex} 

\section{WKB Approximation for Bound States and
Under-Barrier Resonances}

We begin with the standard WKB approximation for a particle
in a 1-dimensional potential well \cite{landau}.
The single-turning-point connection formula for wavefunctions leads to the
Bohr--Sommerfeld quantisation rule for the bound-state energy eigenvalues.

In the original wormhole model \cite{wormhole}, the potential is
\begin{equation}\
V(r) = - \frac{V_0}{(r^{2}+a^{2})^{2}} \,,
\end{equation}
and let $\mu$ denote the reduced mass of the two clusters. By
setting $r=ax$, $V_0=\hbar^{2}a^{2}m^{2}/2\mu$, and writing the energy
as $E= \hbar^{2}\varepsilon / 2\mu a^{2}$ and the stationary states as
$\chi(x)=(x^{2}+1)^{-1 /2}\eta (x)$, we obtain the 1-dimensional, rescaled
Schr\"odinger equation 
\begin{equation}\label{potential}
  -\frac{\d ^{2}\eta}{\d x^{2}}+ \frac{l (l+1)}{x^{2}+1}
  \eta-\frac{m^{2}-1}{(x^{2}+1)^{2}} \eta = \varepsilon \eta \,.
\end{equation}
$l$ is the angular momentum quantum number, and the first $x^2 + 1$
denominator arises from the wormhole metric. In \cite{wormhole} the choice
$m=9$ was made, so the effective potential is
\begin{equation}\label{effpotential}
v_{\rm{eff}}(x) = \frac{l (l+1)}{x^{2}+1}-\frac{80}{(x^{2}+1)^{2}} \,.
\end{equation}

For bound states, with $\varepsilon \leqslant 0$, the
Bohr--Sommerfeld quantisation rule is
\begin{equation}\label{bohrnew}
\int_{-x_0}^{x_0}  p(x,\varepsilon) \, \d x
=\left(n+\frac{1}{2} \right) \pi \,,  
\end{equation}
where $p(x, \varepsilon)= \sqrt{\varepsilon -v_{\mathrm{eff}}(x) }$ is
the `particle' momentum and $-x_0 \leqslant x \leqslant x_0$
denotes the classically accessible region. Explicitly, the turning points
$x=\pm x_0$, where $v_{\mathrm{eff}}(x)=\varepsilon$, are
\begin{equation}
  x_0=\left(\frac{l (l+1) - \sqrt{l^{2}(l+1)^{2}-320\varepsilon}
      -2\varepsilon}{2\varepsilon} \right)^{\tfrac{1}{2}} \,.
\label{turning}
\end{equation}
$n$ is a non-negative integer, equal to the number of nodes of the
wavefunction.

The solutions $\varepsilon_{n,l}$ of (\ref{bohrnew}) have been found
by varying $\varepsilon$ between the (negative) minimum of
$v_{\mathrm{eff}}$ and $0$ for each $l$, and applying linear interpolation
to attain an uncertainty $\Delta \varepsilon_{n,l}<0.01$. The resulting
bound-state energy spectrum is shown in Table 1. $l$ increases in
steps of 2 for given $n$, and the highest state has $l = 8-n$. There are no
resonances for $l \leqslant 8$.

\begin{table}[H]\centering
  \caption{The energy spectrum of the original wormhole model (with
    $m=9$), comparing the WKB approximation and numerical data.}\medskip\small
  \begin{tabular}{ccc}
Band & Numerical Data & WKB Approximation \\ [1mm]
{0-phonon}& [-68.41, -62.67, -49.28, -28.33, 0] &
[-67.91, -62.16, -48.76, -27.77, 0.66] \\ [1mm]
1-phonon& [-45.70, -37.05, -21.65, 0]
& [-45.41, -36.65, -21.23, 0.48]  \\ [1mm]
2-phonon& [-30.76, -26.17, -15.67, 0]
& [-30.47, -25.89, -15.38, 0.325]  \\ [1mm]
3-phonon& [-16.86, -10.51, 0]& [-16.68, -10.34, 0.20]  \\ [1mm]
4-phonon& [-9.36, -6.31, 0]& [-9.26, -6.22, 0.10] \\ [1mm]
5-phonon& [-3.15, 0]& [-3.12, 0.04]  \\ [1mm]
6-phonon& [-1.11, 0]& [-1.12, 0] \\ [1mm]
7-phonon& [0]& [-0.01]  \\ [1mm]
  \end{tabular}
  \end{table}
The WKB results align closely with the numerical ones. The errors are
generally less than $2\%$, and the absolute errors decrease as the
phonon number increases. Fig. 2 shows that the WKB approximation is a
great improvement on the anharmonic oscillator approximation used in
\cite{wormhole}. 
\begin{figure}[H] 
  \centering 
  \input{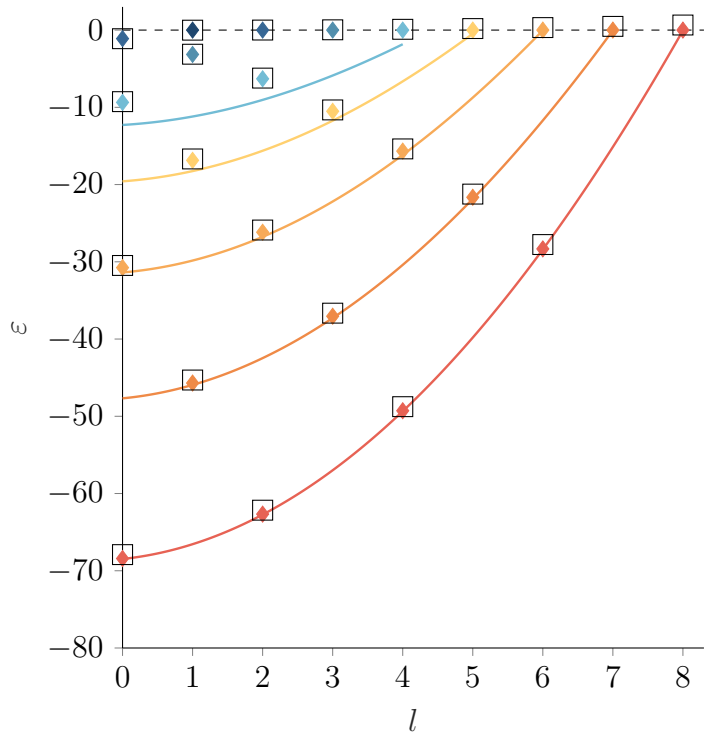}
      \captionsetup{justification=centering}
  \caption{Numerical bound states (dots), the anharmonic oscillator
    approximation (curves) and the WKB approximation (boxes), for the
    original wormhole model. $n$ increases from 0 (red) to 4 (light blue)
    and beyond.}
\end{figure} 

\vspace{1em plus 0.5ex minus 0.2ex} 

We now introduce Shepard's WKB approximations
\cite{shepard} for calculating the complex energies of under-barrier
resonances in a potential $v(x)$. Shepard delineates two versions: the
`simple WKB approximation' and the `improved WKB approximation'.
The latter includes modifications when the single-turning-point
connection formula is inadequate. We first use the simple WKB
approximation, an extension of the standard WKB method.

A resonance corresponds to a pole in the scattering matrix at a
complex energy
\begin{equation}
\varepsilon=\varepsilon_r - \half i \gamma \,,
\end{equation}
with $\varepsilon_r$ the real energy and $\gamma$ the width. 

\begin{figure}[H]
  \centering
  \includegraphics[width=11cm]{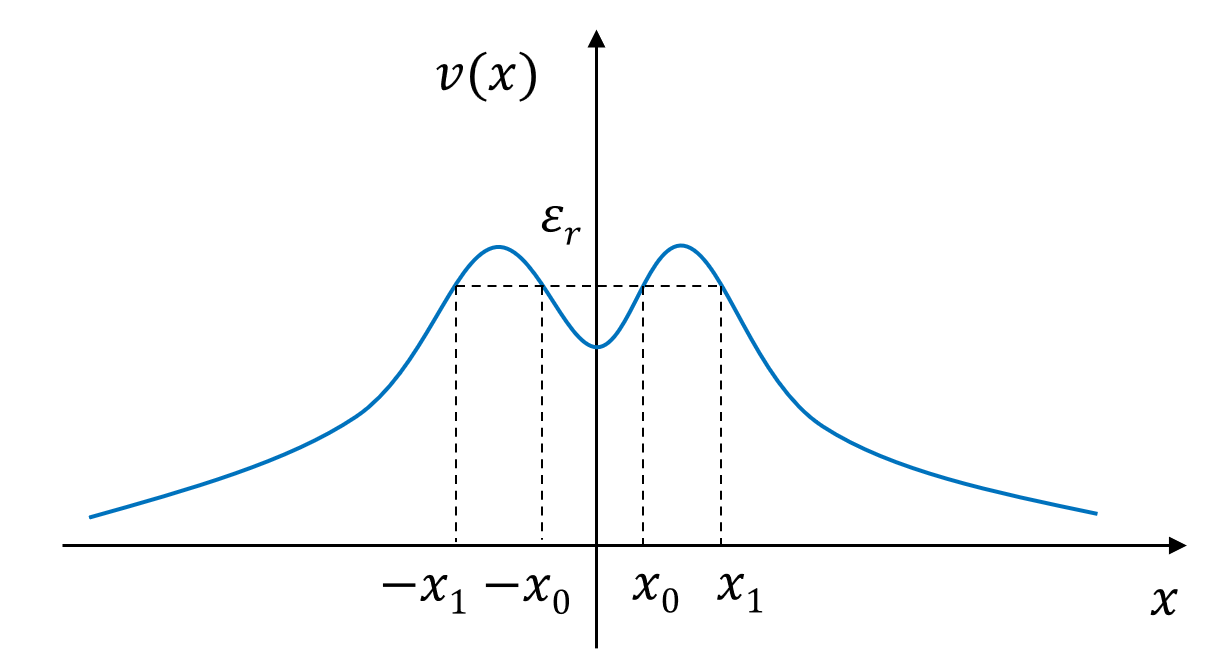}
  \captionsetup{}
  \caption{The turning points $\pm x_0$ and $\pm x_1$, for real energy
 $\varepsilon_r$ (horizontal, dashed line) in double-humped potential $v(x)$.}
\end{figure}

Shepard's approach is particularly suited for a particle resonance
in a double-humped potential well (see Fig. 3), with $\gamma$ significantly
smaller than $\varepsilon_r$. $\pm x_0$ and $\pm x_1$ are defined
as the locations where $v(x) = \varepsilon_r$. By imposing purely-outgoing
wave boundary conditions and applying the reversible connection
formula within the standard WKB framework \cite{langer,miller}, one derives
formulae for $\varepsilon_r$ and $\gamma$. Firstly, we define the integral
\begin{equation}\label{w_1}
  W_1(\varepsilon)= \int_{-x_0}^{x_0} p (x,\varepsilon) \, \d x \,.
\end{equation}
Then, as in (\ref{bohrnew}), the real resonance energies $\varepsilon_r$
satisfy the Bohr--Sommerfeld quantisation rule
\begin{equation}\label{realenergy}
  W_1(\varepsilon_r)=\left(n +\frac{1}{2}\right)\pi
\end{equation}
for $n \in \mathbb{N}$, and $n$ is the number of nodes of the
wavefunction between $-x_0$ and $x_0$. The width $\gamma$ corresponding
to $\varepsilon_r$ is then given by
\begin{equation}\label{width}
  \gamma (\varepsilon_r)= 2 \, \nu (\varepsilon_r)
  \ln \Big(1+ e^{-2W_2 (\varepsilon_r)}+
  \frac{1}{2}e^{-4 W_2 (\varepsilon_r)} \Big) \,,
\end{equation}
where 
\begin{equation}\label{integral1}
  \nu (\varepsilon_r)
  = \left[ \int_{-x_0}^{x_0} \frac{ \d x}{p(x,\varepsilon_r)} \right]^{-1}
\end{equation}
is the inverse of the classical period $T(\varepsilon_r)$ of the particle in
the well at energy $\varepsilon_r$, and
\begin{equation}\label{integral2}
  W_2 (\varepsilon_r)=\int_{x_0}^{x_1} \left\vert p (x,\varepsilon_r)
  \right\vert \d x \,.
\end{equation}
In the interval $(x_0,x_1)$, $\varepsilon_r$ is less than $v(x)$,
so $p(x,\varepsilon_r)$ is imaginary. 

This simple WKB approach relies on the approximate linearity of the
potential $v(x)$ at the turning points $\pm x_0$ and $\pm x_1$, and
the assumption that turning points are well-separated. These
conditions fail for real energies $\varepsilon_r$ near the peak or base of the
potential well. Therefore, Shepard derives an improved WKB
approximation \cite{shepard}, using two-turning-point connection
formulae and the method of comparison equations \cite{miller,berry}.
The real energy $\varepsilon_r$ now satisfies
\begin{equation}\label{integral0}
  W_1(\varepsilon_r) = \left(n +\frac{1}{2}\right)\pi+ \phi (\varepsilon_r) \,,
\end{equation}
and the corresponding width formula is
\begin{equation}\label{width1}
  \gamma (\varepsilon_r)=2 \ln \Big(1+\alpha(\varepsilon_r)^{-1}e^{-2W_2
    (\varepsilon_r)}\Big)\left(T (\varepsilon_r)- 2\left.\frac{\partial
        \phi}{\partial \varepsilon} \right| _{\varepsilon_r}  \right) ^{-1} \,.
\end{equation}
Here $\phi(\varepsilon_r)$ and $\alpha(\varepsilon_r)$ are defined as
\begin{equation}\label{phi}
  \phi(\varepsilon_r)= \arg {\rm Gam} \left(\frac{1}{2}- \frac{iW_2
  (\varepsilon_r)}{\pi}  \right) + \frac{W_2 (\varepsilon_r)}{\pi} \left(\ln
    \frac{W_2(\varepsilon_r)}{\pi} -1 \right)
\end{equation}
and
\begin{equation}
  \alpha (\varepsilon_r)=(2\pi)^{- 1/2}n!
  \left(\frac{e}{n+ 1/2} \right)^{n + 1/2} \,,
\end{equation}
with ${\rm Gam}$ the Gamma function. 

Note that in eqs.(\ref{integral0}) and (\ref{width1}), the
modification $\phi(\varepsilon_r)$ accounts for states near the peak
of the potential, while $\alpha(\varepsilon_r)$ modifies states near
the base. These modifications are compatible, as both revert to
the simple WKB case outside their respective regions of application. One
can easily check that if a state is localised far from the peak, then
$\phi(\varepsilon_r)$ and $\partial \phi / \partial \varepsilon$
approach zero; whereas if a state is far from the base,
$\alpha(\varepsilon_r)$ approaches 1.

To assess the simple and improved WKB approximations,
we calculate the complex energy $\varepsilon = \varepsilon_r
- i \gamma /2$ of the under-barrier resonance with $n=0$ and $l=10$
in the original wormhole model with $m=9$, comparing with
the numerical result. This is the only under-barrier resonance for
$l \leqslant 10$. We seek the slightly different solutions
$\varepsilon_r$ that satisfy the Bohr--Sommerfeld conditions
(\ref{realenergy}) and (\ref{integral0}) between
the minimum and maximum of $v_{\mathrm{eff}}(x)$. We then insert
$\varepsilon_r$ into eqs.(\ref{width}) and (\ref{width1}), where $x_0$
is as in (\ref{turning}) with $\varepsilon$ replaced by
$\varepsilon_r$, and 
\begin{equation}
  x_1=\left(\frac{l (l+1) + \sqrt{l^{2}(l+1)^{2}-320\varepsilon_r}
      - 2\varepsilon_r}{2\varepsilon_r}\right)^{\tfrac{1}{2}} \,.
\end{equation}
The results for the real energy and width are given in Table 2.

\begin{table}[htb]\centering
  \caption{The real energy and width of the $n=0$, $l=10$ resonance
    in the original model computed by WKB approximations and numerical
    method.}\medskip\small
\begin{tabular}{|c|ccc|}
\hline
\rule{0pt}{3ex} & Simple WKB & Improved WKB & Numerical method \\ [1mm] 
\hline\hline
\rule{0pt}{2.8ex} Real energy $\varepsilon_r$ &35.996 &35.530 & 34.863\\ [1mm]
\rule{0pt}{2.8ex} Width $\gamma$ & 0.704 & 0.798 & 0.669 \\ [1mm]
\hline
  \end{tabular}
  \end{table}

The improved WKB approximation yields a more accurate result for
$\varepsilon_r$, but the width $\gamma$ is less accurate, being
highly sensitive to changes in the real energy. Nevertheless,
when we include the Coulomb correction, we will preferentially employ
the improved WKB approximation for the computation of under-barrier
resonance energies and widths.
  
\vspace{1em plus 0.5ex minus 0.2ex} 

\section{WKB Approximation for Over-Barrier Resonances}

As the Coulomb-corrected wormhole model lowers the threshold energy to
4.73 MeV, we expect certain states to become over-barrier resonances.
These correspond to poles in the complex energy plane, well away from
the real axis. Resonances with \( \varepsilon_r > v_{\max} \) have
large widths $\gamma$, so it is not valid to calculate to first order
in \( \gamma / \varepsilon_r \) and separate the conditions on
\( \varepsilon_r \) from those on \( \gamma \). We need a single condition
on the complex energy \( \varepsilon = \varepsilon_r - i \gamma /2 \). 

The improved WKB approximation cannot be directly used, as it requires
the equation \( v(x) =\varepsilon_r \) to have four real solutions
at \( x = \pm x_0 \) and \( x = \pm x_1 \). Instead, it is
necessary to analytically continue this approximation. We begin by defining
the two complex turning points, \(\tilde{x}_{0}(\varepsilon)\)
and \(\tilde{x}_{1}(\varepsilon)\) in the lower complex
$\varepsilon$-plane. These are solutions of \( v(z) = \varepsilon \) and are
required to be the analytic continuation of the real turning points \(
x_0(\varepsilon_r) \) and \( x_1(\varepsilon_r) \) for
\( \varepsilon_r \leqslant  v_{\max} \) (see Fig. 4).

\begin{figure}[H] 
  \centering 
  \input{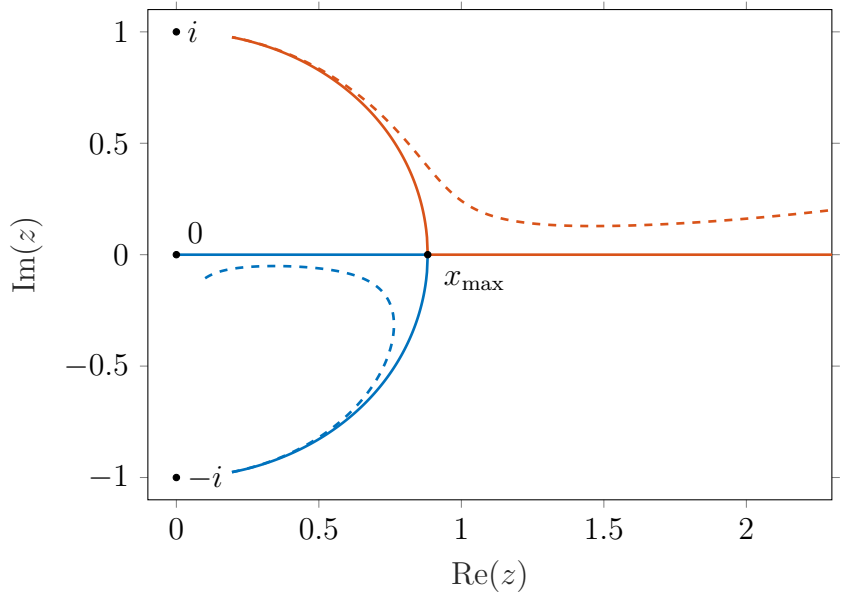}
      \captionsetup{justification=centering}
  \caption{The trajectories of $\tilde{x}_0(\varepsilon_r)$ (blue) and
    $\tilde{x}_1(\varepsilon_r)$ (orange) at fixed widths $\gamma=0$ (solid)
    and $\gamma=3$ (dashed) for the original effective potential
    $v_{\rm eff}$ with $l=9$, $m=9$.}
\end{figure} 

Next, we analytically continue the real integrals (\refeq{w_1}) and
(\refeq{integral2}). Specifically, for $z,\varepsilon \in \mathbb{C}$, the
generalised momentum function is defined as
\( p(z, \varepsilon) = \sqrt{\varepsilon - v(z)} \). We then introduce the
contour integrals, \( \widetilde{W}_{1}(\varepsilon) \) and
\( \widetilde{W}_{2}(\varepsilon) \), for fixed
\( \varepsilon = \varepsilon_r - i \gamma / 2 \),
\begin{equation}\label{w1tilde}
  \widetilde{W}_1(\varepsilon)= \int_{\gamma_1}^{} p(z,\varepsilon) \, \d z \,,
\end{equation}
where \( \gamma_1 \) is a contour from \( -\tilde{x}_0(\varepsilon) \) to \(
\tilde{x}_0(\varepsilon) \), and
\begin{equation}\label{w2tilde}
  \widetilde{W}_2(\varepsilon)=\int_{\gamma_2}^{} ip(z,\varepsilon) \, \d z \,,
\end{equation}
where \( \gamma_2 \) is a contour from \( \tilde{x}_0(\varepsilon) \) to \(
\tilde{x}_1(\varepsilon) \). \( p(z, \varepsilon) \) is double-valued,
necessitating the choice of a branch (see Fig. 5). We set the branch cut
\(\mathcal{C}_{p}\) to be
$$
\left\{ z \in \mathbb{C}_{\infty}: \varepsilon-v(z) \in [-\infty,0] \right\} \,,
$$
and select \( p(0, \varepsilon) \) so that \( \arg(\varepsilon - v(0))
\in (-\pi, \pi] \), ensuring that \( \widetilde{W}_{1}(\varepsilon) \)
and \( \widetilde{W}_{2}(\varepsilon) \)
are the analytic continuations of the real integrals.

\begin{figure}[H]
  \centering
  \includegraphics[width=9cm]{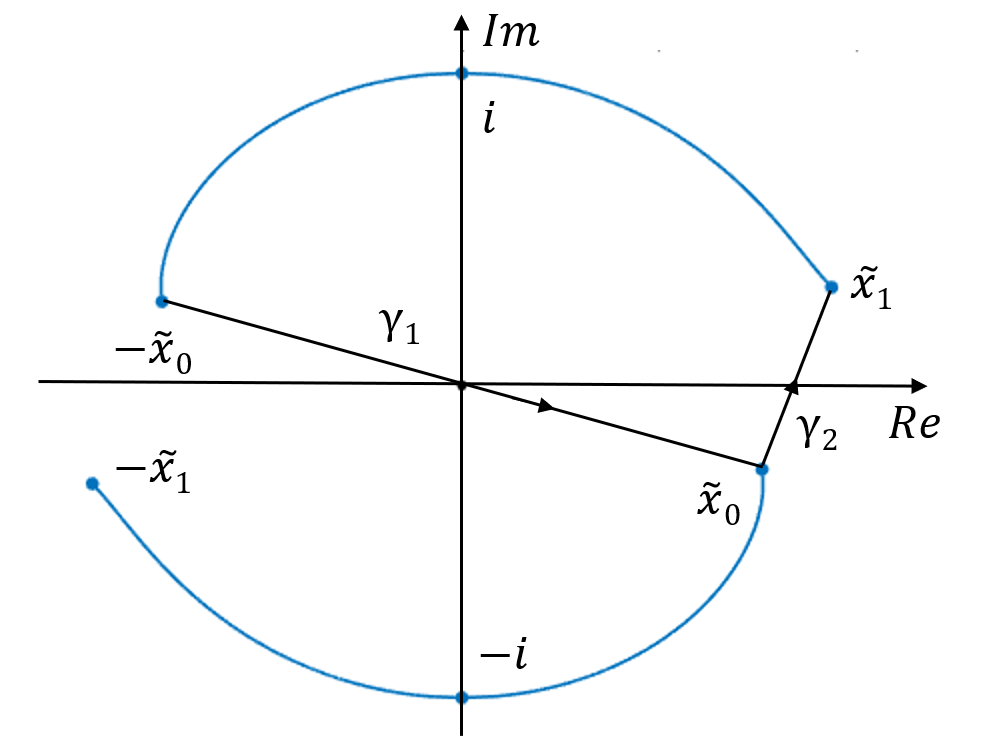}
  \captionsetup{}
  \caption{The branch cut $\mathcal{C}_{p}$ (blue), illustrated for
    the function $p(z,\varepsilon)$ with $\varepsilon_r=27.313$ and
    $\gamma=3$, and $v_{\rm eff}$ as in Fig. 4.}
\end{figure}

Finally, the condition for the complex resonance energy \( \varepsilon \)
becomes
\begin{equation}\label{res}
  \widetilde{W}_1(\varepsilon)=\left(n+\frac{1}{2} \right) \pi
  + \widetilde{\phi}(\varepsilon)
\end{equation}
for \( n \in \mathbb{N} \), where $\widetilde{\phi}$ is defined as
\begin{equation}\label{res2}
 \widetilde{\phi}(\varepsilon) = \frac{\widetilde{W}_2(\varepsilon)}{\pi}
 \left(\ln \frac{\widetilde{W}_2(\varepsilon)}{\pi} -1 \right)
 - \frac{i}{2}  \ln \left(\frac{{\rm Gam} \left(\frac{1}{2}
       - i\frac{\widetilde{W}_2(\varepsilon)}{\pi}\right)
     \left(1+ e^{-2 \widetilde{W}_2(\varepsilon)} \right)}
 {{\rm Gam} \left(\frac{1}{2}
  + i\frac{\widetilde{W}_2(\varepsilon)}{\pi} \right) }  \right) \,.
\end{equation}
This is a consistent generalization of results in ref.\cite{mur}, which
examines over-barrier resonances for single-humped potentials.

As a test, we now apply this method to  $v_{\rm eff}$, the effective potential
(\refeq{effpotential}) of the original wormhole model with $m=9$ and
varying $l$. We find several over-barrier resonances by searching
for complex energies $\varepsilon$ that satisfy condition
(\refeq{res}) near the top of the barrier, and compare with numerical
data. The findings for the lowest-energy resonances are presented in Table 3.

\begin{table}[htb]\centering
  \caption{The scaled real energies $\varepsilon_{n,l}$ and
    widths $\gamma$ of the lowest-energy, over-barrier
    resonances with $5 \leqslant l \leqslant 9$ and $n = 10 - l$
    in the original wormhole model, computed by solving the over-barrier WKB
    equation (\refeq{res}), and by a numerical method.}\medskip\small
  \begin{tabular}{|c|c|ccccc|}
    \hline
    \rule{0pt}{3ex}  & Methods & $l=5$ & $l=6$ & $l=7$ & $l=8$ & $l=9$
    \\ [1mm]
    \hline\hline
\rule{0pt}{3ex} Barrier Height & N/A & 2.81 & 5.51 & 9.80 & 16.20 &
25.31 \\ [1mm]
\hline\hline
\multirow{ 2}{*}{Real Energy} & \rule{0pt}{2.8ex} WKB &$2.91$ & $6.59$
&  $11.77$ & $18.36$ & $26.30$ \\ [1mm]    
& \rule{0pt}{2.8ex} Numerical & 2.99 & 6.63 & 11.72 & 18.18 & 25.94 \\ [1mm]
\hline\hline
\multirow{ 2}{*}{Width} & \rule{0pt}{2.8ex} WKB &5.19 & 6.48  &  6.42
& 5.01 & 2.77 \\ [1mm]    
& \rule{0pt}{2.8ex} Numerical & 5.08 & 6.26 & 6.13 & 4.73 & 2.58 \\ [1mm]
\hline
  \end{tabular}
\end{table}
From the barrier heights, we verify that these
resonances are all over-barrier. The accuracy achieved for the widths is
notable, given their sensitivity to small errors in the real energies.

\vspace{1em plus 0.5ex minus 0.2ex} 

\section{The Coulomb Correction}

In this section we incorporate the Coulomb potential in the wormhole model;
this is essential for addressing the threshold energy issue and
analysing resonances. We retain the wormhole metric (\ref{metric}) but
introduce the modified potential (\refeq{Pot}).

With analogous rescaling and the same substitution
$\chi (x)=(x^{2}+1)^{-1 /2}\eta (x)$ as in (\refeq{potential}), we
derive the 1-dimensional Schrödinger equation
\begin{equation}\label{potential1}
  -\frac{\d ^{2}\eta}{\d x^{2}}+ \left(\frac{l (l+1)}{x^{2}+1}+
    \frac{1}{(x^{2}+1)^{2}}  -
    \frac{m^{2}\sigma^{4}}{(x^{2}+\sigma^{2})^{2}}  + \frac{C_0
      \beta}{(x^{\kappa}+ \beta ^{\kappa})^{1 / \kappa}} \right) \eta
  =\varepsilon \eta \,,
\end{equation}
where
\( b = a \beta \), \( d = a \sigma \), \( Q_0 = \hbar^{2} C_0 \beta /
2\mu a \), \( V_0 = \hbar^{2} a^{2} m^{2} \sigma^{4} / 2\mu \) and \(
E = \hbar^{2} \varepsilon / 2 \mu a^{2} \). The reduced mass of the clusters
is $\mu=2982$ MeV. The effective potential \( u_{\mathrm{eff}}(x) \) is now
\begin{equation}
u_{\mathrm{eff}}(x)=\frac{l (l+1)}{x^{2}+1}+ \frac{1}{(x^{2}+1)^{2}}
- \frac{m^{2}\sigma^{4}}{(x^{2}+\sigma^{2})^{2}}  + \frac{C_0
  \beta}{(x^{\kappa}+ \beta ^{\kappa})^{1 / \kappa}} \,,
\end{equation}
and we refer to the last term as the dimensionless
Coulomb potential \( u_{\rm{Cou}}(x) \). Fig. 6 shows plots of \(
u_{\rm{Cou}}(x) \) for fixed \( \beta \) and \( C_0 \), and varying
\( \kappa \), and Fig. 7 compares $u_{\rm eff}$ with the original
$v_{\rm eff}$.

\begin{figure}[H] 
  \centering 
  \input{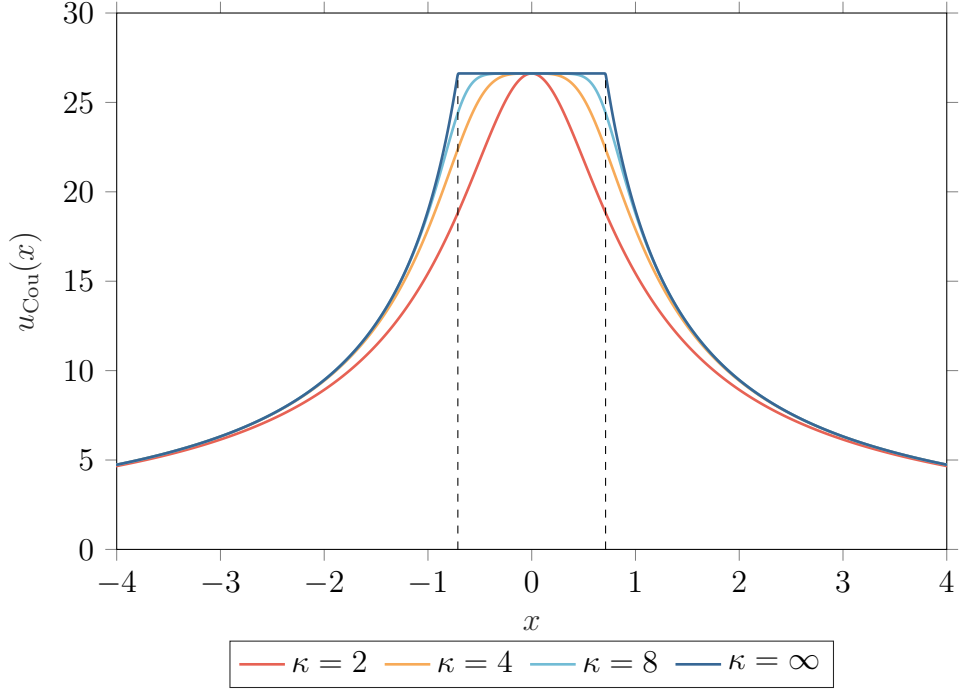}
      \captionsetup{justification=centering}
  \caption{The dimensionless Coulomb potential $u_{\rm{Cou}}(x)$
    for various $\kappa$ but fixed $\beta=0.7119$ and
    $C_0=26.62$. Dotted lines: $x=\pm \beta$.}
\end{figure} 

As \( \kappa \longrightarrow \infty \), $u_{\rm{Cou}}(x)$
becomes constant within the interval \( x \in [-\beta,
\beta] \) and equals \( C_0 \beta / x \) outside, resembling the
potential created by a charge uniformly distributed on a spherical
surface. This type of potential is widely used in models of alpha
decay \cite{griffiths,bertulani}. The constant
behaviour for \( x \in [-\beta, \beta] \) helps
preserve the real energy spectrum of the original model which already
fits experimental data well, as the wavefunctions of the states are
predominantly localised in the interval \( x \in [-1, 1] \). The only
significant change is an overall energy shift, but this disappears
when we calculate energies relative to the ground state. However, 
for $\kappa = \infty$, the Coulomb potential is non-differentiable.
Therefore, we set \( \kappa = 8 \) from here on, ensuring the
smoothness of the potential while retaining the advantageous
properties of the limiting case.

\begin{figure}[H] 
  \centering 
  \input{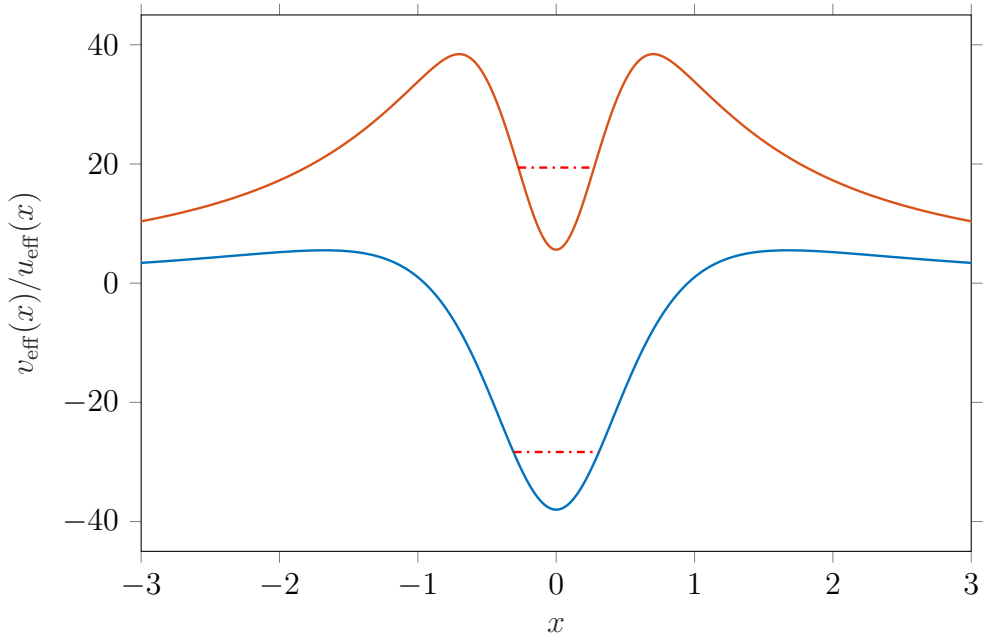}
      \captionsetup{justification=centering}
  \caption{The effective potentials $v_{\rm eff}(x)$ in the original
    wormhole model (blue) and $u_{\rm eff}(x)$ in the Coulomb-corrected
    model (orange), for $l=6$. The red dash-dotted lines show the
    real energies of 0-phonon, $l=6$ states. The Coulomb potential shifts the
    energy upwards and converts the bound state into a resonance.}
\end{figure} 

In Sect. 2 we have adopted $m=8$. To isolate the effect of the Coulomb
interaction, we first set the Coulomb potential $u_{\mathrm{Cou}}(x)$
to zero, thereby recovering the original wormhole model except for the
values of $m$ and $\sigma$. We then determine $\sigma$ and the length
scale $a$ by an optimization process. For each value of $\sigma$, we
perform a least-squares fit to the real energies $E$ of the
\( l =0, 2, 4, 6, 8 \) states in the 0-phonon band. This yields an energy
calibration factor $\hbar^2/2\mu a^2$ relating $E$ and $\varepsilon$, and
the corresponding relative residual; minimizing that residual gives
$\sigma=0.6652$. However, recall from \cite{wormhole} that, by
considering the kinetic energy term, one finds the relationship
between the physical cluster separation $R$ and the dimensionless
coordinate $x$,
\begin{equation}\label{separation}
  R^{2}=\left(x^{2}+\frac{5}{11} \right) a^{2} \,.
\end{equation}
Therefore, we have preferred to set $\sigma = \sqrt{5/11} \approx
0.6742$, a choice that follows naturally from assuming the strong
interaction is inversely proportional to the fourth power of the
separation $R$. For this choice, the calibration factor becomes
$\hbar^2/2\mu a^2=0.2265$ MeV and the wormhole radius $a=5.37$ fm,
which are close to those in the original model
\cite{wormhole}. Additionally, the threshold energy (without Coulomb
repulsion) is reduced from the original value 15.19 MeV to 10.76 MeV
due to setting $m=8$.

We next turn on the Coulomb potential $u_{\mathrm{Cou}}(x)$,
characterised by its parameters $C_0$ and $\beta$. The
Coulomb potential does not significantly perturb the 0-phonon
spectrum for $\kappa=8$ as we argued above, so the previously fixed
$\sigma$ and $a$ remain valid. By considering the behaviour of
$u_{\mathrm{Cou}}(x)$ around $x=0$, we find that the Coulomb repulsion
shifts the threshold energy by $Q_0/b=\hbar ^{2} C_0 / 2 a^{2}$ as
shown in Fig. 7. Imposing the experimental threshold
$E_{\mathrm{thresh}}=4.73$ MeV determines the shift $(10.76-4.73) \,
\mathrm{MeV}=6.03$ MeV and thus fixes the parameter $C_0=26.62$.

Finally, the value of $\beta$ is determined by requiring the
physical Coulomb tail behaviour
\begin{equation} \label{infcons}
V_{\mathrm{Cou}}(r)=\frac{Q_0}{r}+O \left(\frac{1}{r^{2}} \right)
\sim \frac{16 e^{2}}{4 \pi \epsilon_0 r} \,,
\end{equation}
where $16e ^{2}$ is the product of the electric charges of the alpha particle and Oxygen-16. Rearranging \eqref{infcons} we obtain
\begin{equation}
C_0 \beta= \frac{8 \mu a e^{2}}{\pi \epsilon_0 \hbar ^{2}}
=16 \alpha \sqrt{2 \mu \, \frac{2 \mu a^{2}}{\hbar ^{2}}  }=18.95 \,,
\end{equation}
where $\alpha \approx 1/137$ is the fine structure constant, and in
the last equality, we set $\mu=2982$ MeV and the energy
calibration factor $\hbar  ^{2} /2 \mu a^{2} =0.2265$ MeV. This yields
$\beta=0.7119$ and completes the determination of all parameters
in the Coulomb-corrected wormhole model. We can now generate
quantitative predictions for the Neon-20 spectrum.

\vspace{1em plus 0.5ex minus 0.2ex} 

\section{Energy Spectrum and Comparison with Observed Neon-20 States}

Here we present the Coulomb-corrected spectrum of real energies and
widths of states in the 0-, 1-, and 2-phonon bands up to $l=10$,
using the appropriate WKB methods discussed earlier. To obtain a
physical, real energy value, we subtract from the dimensionless energy
$\varepsilon_{n,l}$ the ground-state energy
\( \varepsilon_{0,0} \) and multiply by the calibration
factor 0.2265 MeV. The same factor relates $\gamma$ to the physical
width $\Gamma$. The effective potential barrier height is
considerably higher than the threshold energy 4.73 MeV for all
angular momenta $l$.

There are bound states, under-barrier and over-barrier resonances.
We use the WKB method for bound states to compute the energies
for \( n = 0 \) and \( l = 0, 2, 4 \). For states with
\( n = 0 \), \( l = 6, 8, 10 \) and \( n = 1 \), \( l = 1, 3, 5 \), we
employ the improved WKB method for under-barrier resonances, and for states
with \( n = 1 \), \( l = 7,9 \) and \( n = 2 \), \( l = 0, 2, 4, 6 , 8\),
we apply the over-barrier WKB method. The under- and over-barrier
resonances are interleaved in energy, because they have different $l$ values.

\begin{table}[H]\centering
\caption{Comparison between experimental and Coulomb-corrected
energy levels of Neon-20. $n$;$l$ denote phonon number and angular
momentum. BS, UBR and OBR denote bound state, under-barrier
resonance and over-barrier resonance, respectively. The brackets
indicate there are other potential candidates for the
identification.}\medskip\small
  
\begin{tabular}{|c|c|c|c|c|c|c|}
    \hline
    \multicolumn{3}{|c|}{Experiment} &
\multicolumn{4}{c|}{Model}\rule{0pt}{3ex} \\ [1mm]
    \hline
    \rule{0pt}{3ex} $E_r$ (MeV)  &  $\Gamma$ (keV) & $J^{\pi}$& $E_r$
(MeV) &$\Gamma$ (keV) & $n$;$l$ & Type \\ [1mm]
\hline\hline
\rule{0pt}{2.8ex} 0 &0 & $0^{+}$&0 & 0 & 0;0 & BS\\ [1mm]
\rule{0pt}{2.8ex} 1.63 & 0 & $2^{+}$& 1.31 & 0 & 0;2 & BS\\ [1mm]
\rule{0pt}{2.8ex} 4.25 & 0 & $4^{+}$& 4.38 & 0 & 0;4 & BS\\ [1mm]
\rule{0pt}{2.8ex} 5.79 & $0.028$ & $1^{-}$& 5.73 & 0.0015 & 1;1 & UBR\\ [1mm]
\rule{0pt}{2.8ex} 7.16 & 8.2 &$3^{-}$& 7.63 & 16 & 1;3 & UBR\\ [1mm]
\rule{0pt}{2.8ex} $\approx 8.7 $ & $\approx 800$ & $0^{+}$& 7.82 & 850
& 2;0 & OBR\\ [1mm]
\rule{0pt}{2.8ex} 8.78 & 0.11 & $6^{+}$& 9.12 & 0.19 & 0;6 & UBR\\ [1mm]
\rule{0pt}{2.8ex} 9.00 & $\approx 800$ & $2^{+}$& 8.68 & 1200 & 2;2 &
OBR\\ [1mm]
\rule{0pt}{2.8ex} 10.26 & 150 & $5^{-}$& 11.00 & 320 & 1;5 & UBR\\ [1mm]
\rule{0pt}{2.8ex} 10.80 & 350 & $4^{+}$& 10.84 & 2300 & 2;4 & OBR\\ [1mm]
\rule{0pt}{2.8ex} (15.16) & 60 &$6^{+}$& 14.55 & 4500 & 2;6 & OBR\\ [1mm]
\rule{0pt}{2.8ex} 15.37 & 110 & $7^{-}$& 15.94 & 1000 & 1;7 & OBR\\ [1mm]
\rule{0pt}{2.8ex} 15.87 & 100 & $8^{+}$& 15.58 & 38 & 0;8 & UBR\\ [1mm]
\rule{0pt}{2.8ex}  &  &  & 21.65 & 7100 & 2;8 & OBR\\ [1mm]
\rule{0pt}{2.8ex} 22.80 & 500 & $9^{-}$ & 22.54 & 2200 & 1;9 & OBR\\ [1mm]
\rule{0pt}{2.8ex} (25.67) & $\approx 400$ &  & 23.72 & 330 & 0;10 & UBR\\ [1mm]
\rule{0pt}{2.8ex}  &  &  & 29.19 & 8900 & 2;10 & OBR\\ [1mm]
\hline
  \end{tabular}
  \end{table}

Table 4 compares the computed results with experimental data from the
TUNL \cite{tunl} and ENSDF \cite{ensdf} tables for Neon-20. The Coulomb-corrected model retains the general pattern of real energies in
the original model, but we have now also estimated the widths of the
states above threshold. The real energies $E_r$ agree well with
experimental data. Rotational bands constructed from these states
have been identified in \cite{wormhole}, following refs.
\cite{bouten, buck, michel, fuji, cseh, marevic} and the recent
discussion by Bijker and Iachello \cite{bijker}. Fig. 8 presents a
real energy spectrum diagram, showing the $0^+_1$, $0^-_1$
and $0^+_4$ rotational bands. Fig. 9 shows the widths. 

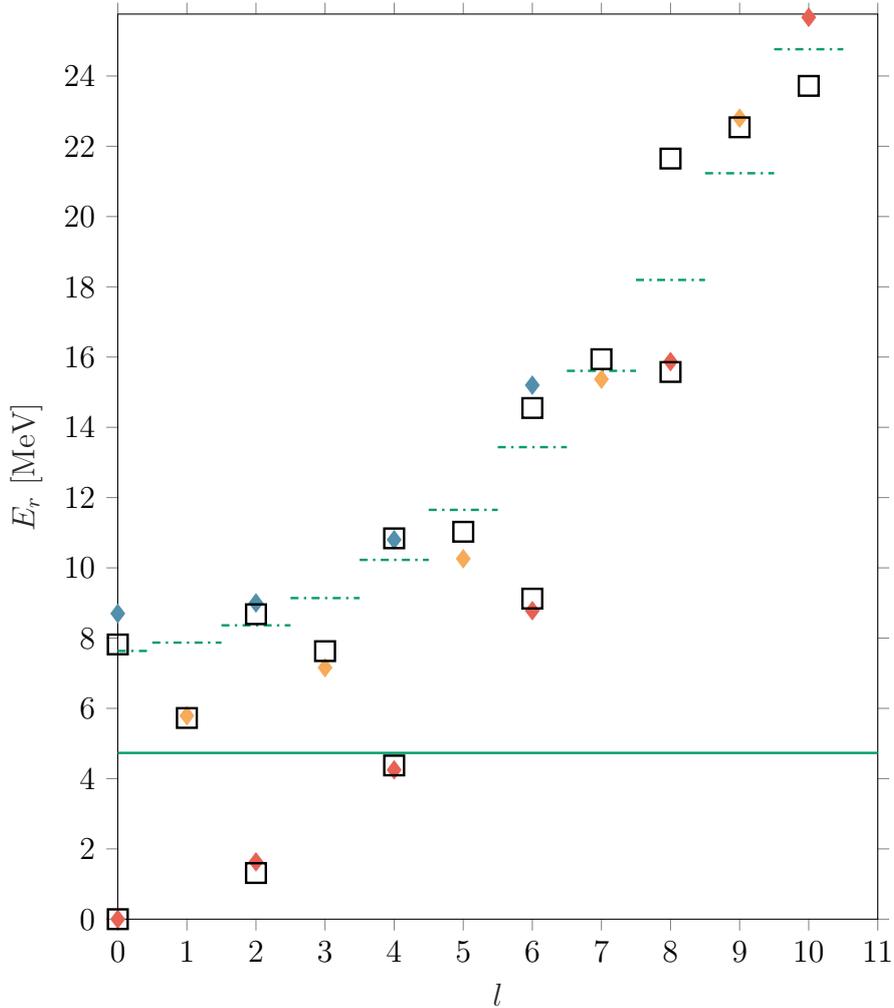
\begin{figure}[H] 
  \centering 
  \input{energylevels.tex}
      \captionsetup{justification=centering}
  \caption{Experimental real energies (dots), WKB-approximated energies in
    Coulomb-corrected model (black boxes). Green solid line: threshold
    energy; Green dash-dot lines: effective potential barrier height for
    each $l$.}
\end{figure} 

The experimental \( 0_{1}^{+} \) rotational band is identified with
our model's 0-phonon band. The \( l = 0 \) state is the stable Neon-20 ground
state; the \( l = 2, 4 \) states are also below threshold and
undergo \(\gamma\)-decay rather than \(\alpha\)-decay. The states
with \( l = 6, 8 \) have calculated widths comparable to those observed
experimentally, reinforcing our identification of the 15.87 MeV
\( 8^{+} \) state as belonging to the \( 0_{1}^{+} \) band. Our model
predicts a further \( l = 10 \) under-barrier resonance with
\( E_{r} = 23.72 \) MeV and \( \Gamma = 330 \) keV in
this band; its energy is lower than that of the similar state in the
original model (see Table 3). An observed state of high but
undetermined angular momentum, with energy 25.67 MeV and width
around 400 keV, may be a suitable candidate to match this predicted
$l=10$ state, but another potential candidate is an observed
$10^+$ state at 27.50 MeV with unknown width and apparently unconfirmed 
alpha decay.

The well-established \( 0^-_1 \) rotational band with states from
\( l = 1 \) up to \( l = 7 \) is identified with the model's 1-phonon band.
The general agreement between the width predictions and experimental
data supports the identification of the \( 7^{-} \) resonance
at 15.37 MeV as belonging to this band. The model predicts
a $9^-$ resonance in the 1-phonon band with $E_r=22.54$ MeV and
$\Gamma=2200$ keV. We match this with the observed
state at $22.80$ MeV and (smaller) width $330$ keV, in agreement with its
assignment to the $0_{1}^{-}$ band in \cite{fuji}. The \( 0_{4}^{+} \)
rotational band is recognised as the model's higher-nodal, 2-phonon band.

\begin{figure}[H] 
\centering 
\input{widths.tex}
\captionsetup{justification=centering}
\caption{The widths of states in the 0-phonon band (red), 1-phonon
band (blue) and 2-phonon band (orange). Solid lines connect the
model's predictions; dotted lines connect experimental data points.}
\end{figure}
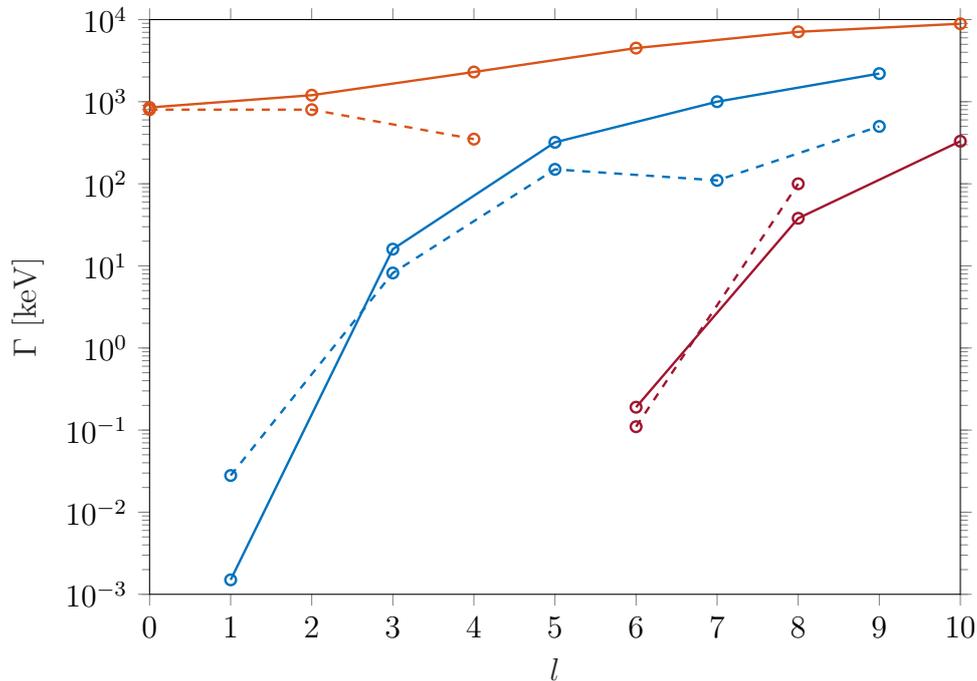 

Fig. 9 shows a logarithmic plot of the widths $\Gamma$ of resonances
in the various bands. Widths larger than 800 keV, especially those in
the $0_{4}^{+}$ band are explained by these resonances being over-barrier. The width discrepancy for the \( n = 1, l = 1 \) state is somewhat expected,
as the state is close to the 4.73 MeV threshold, making its width
highly sensitive to small changes in its real energy. The discrepancy
for the \( n = 2, l = 4 \) (and probably $l=6$) states is more
disappointing. In our model, the widths are observed to increase more
rapidly with $l$ than in the experimental data, indicating that the
potential barriers for large \( l \) should be wider or higher. An
alternative interpretation is that the $6^+$ state has been
misidentified; there is another \( 6^{+} \) state in the 14 -- 16 MeV
region which could be assigned to the $0_4^{+}$ band, as suggested by
Michel et al. \cite{michel}. This state has energy 15.35 MeV but
no width data.

According to our model, states in the 3-phonon and higher-phonon bands
are predicted to have widths exceeding 3000 keV, too large to align
with any experimental observation. 

\vspace{1em plus 0.5ex minus 0.2ex} 

\section{Conclusions}

We have extended the original Neon-20 wormhole model \cite{wormhole}
describing the two-cluster rovibrational dynamics of an alpha particle
and Oxygen-16, by adding a Coulomb potential. The combined
wormhole geometry, short-range nuclear attraction and Coulomb
repulsion gives a consistent dynamical picture of the clusters.
The Coulomb repulsion lowers the threshold energy for unbinding of
the alpha particle, enabling a correct distinction between bound
states and above-threshold resonances, as well as calculations of
the resonance widths.

In detail, we have analysed the 0-, 1- and 2-phonon excitations of the
lowest-frequency vibrational mode of the
five-alpha-particle bipyramid. The rovibrational bands are recognised
as the experimental $0^+_1$, $0^-_1$ and $0^+_4$ bands, the last of
these being the higher-nodal band of Fujiwara et al. \cite{fuji}. 
To compute both energies and widths, we have used the standard WKB
approximation for bound states, and Shepard's improved WKB approximation
\cite{shepard} for under-barrier resonances.
Additionally, for over-barrier resonances we have generalized
Shepard's approach by analytically continuing the improved WKB
approximation. These WKB methods outperform the anharmonic oscillator
approximation, previously applied in ref.\cite{wormhole}.

The energy calibration factor and the radius of the wormhole's throat remain
close to those in the original model, resulting in a separation of
3.61 fm between the alpha particle and Oxygen-16 cluster in the
bipyramid, consistent with calculations of Zhou et al. \cite{zhou}.
With the inclusion of the Coulomb potential, the model achieves real energies
within 8\% of observed values. More importantly, the Coulomb-corrected
model provides a reasonable fit for the widths of states. In
particular, the large widths in the 2-phonon band are explained
by the states being over-barrier resonances, as suggested in
\cite{fuji}. The model supports the identification of the 15.87 MeV
\( 8^{+} \) resonance and the 15.37 MeV \( 7^{-} \) resonance as belonging to,
respectively, the \( 0^+_1 \) and \( 0^-_1 \) bands. Additionally, it
predicts a $9^{-}$ resonance at $E_{r}=22.54$ MeV with
$\Gamma=2200$ keV, lending further support to the assignment of the
observed 22.80 MeV state to the $0^{-}_{1}$ band, although the
predicted width is too large. The model also predicts a possible
\( 10^{+} \) under-barrier resonance with energy 23.87 MeV
and width 310 keV in the $0^+_1$ band. Finally, the too-large
predicted width of the 2-phonon, $6^+$ state suggests an alternative
experimental candidate for this.

In future, Coulomb corrections could be added to the dynamics of other
vibrational modes of the Neon-20 bipyramid \cite{bijker}, in particular the
back-to-back dynamics leading to a symmetric splitting into Carbon-12 and
two alpha particles. Applications of our methods to the excitation
spectra of other two-cluster nuclei are also likely to be worthwhile.

\section*{Acknowledgement}

RB thanks St John's College, Cambridge for funding this work as a
summer project. NSM is partially supported by STFC consolidated
grant ST/P000681/1.

\end{document}

%% file: energylevels.tex
%
%
\definecolor{mycolor1}{rgb}{0.90588,0.38431,0.32941}%
\definecolor{mycolor2}{rgb}{0.96863,0.66667,0.34510}%
\definecolor{mycolor3}{rgb}{0.32157,0.56078,0.67843}%
\definecolor{mycolor4}{rgb}{0.01569,0.61569,0.41961}%
\begin{tikzpicture}

\begin{axis}[%
  width=10cm,
  height=12 cm,
at={(0.758in,0.492in)},
scale only axis,
xmin=0,
xmax=11,
tick align=outside,
xlabel style={font=\color{white!15!black}},
xlabel={$l$},
ymin=0,
ymax=25.7622656455279,
ylabel style={font=\color{white!15!black}},
ylabel={$E_{r}$ [MeV]},
axis background/.style={fill=white}
]
\addplot[only marks,line width=0.9pt, mark=diamond*, mark options={}, mark size=3pt, color=mycolor1, fill=mycolor1, forget plot] table[row sep=crcr]{%
x	y\\
0	0\\
2	1.63\\
4	4.25\\
6	8.78\\
8	15.87\\
10	25.67\\
};
\addplot[only marks,line width=0.9pt,  mark=diamond*, mark options={}, mark size=3pt, color=mycolor2, fill=mycolor2, forget plot] table[row sep=crcr]{%
x	y\\
1	5.79\\
3	7.16\\
5	10.26\\
7	15.37\\
9	22.8\\
};
\addplot[only marks,line width=0.9pt,  mark=diamond*, mark options={}, mark size=3pt, color=mycolor3, fill=mycolor3, forget plot] table[row sep=crcr]{%
x	y\\
0	8.7\\
2	9\\
4	10.8\\
6	15.2\\
};
\addplot[only marks,line width=0.9pt,  mark=square, mark options={}, mark size=3.7pt, draw=black, forget plot] table[row sep=crcr]{%
x	y\\
0	0\\
};
\addplot[only marks,line width=0.9pt,  mark=square, mark options={}, mark size=3.7pt, draw=black, forget plot] table[row sep=crcr]{%
x	y\\
1	5.72648799618711\\
};
\addplot[only marks,line width=0.9pt,  mark=square, mark options={}, mark size=3.7pt, draw=black, forget plot] table[row sep=crcr]{%
x	y\\
2	1.31339808828965\\
};
\addplot[only marks,line width=0.9pt,  mark=square, mark options={}, mark size=3.7pt, draw=black, forget plot] table[row sep=crcr]{%
x	y\\
3	7.62921447075495\\
};
\addplot[only marks,line width=0.9pt,  mark=square, mark options={}, mark size=3.7pt, draw=black, forget plot] table[row sep=crcr]{%
x	y\\
4	4.37519821847404\\
};
\addplot[only marks,line width=0.9pt,  mark=square, mark options={}, mark size=3.7pt, draw=black, forget plot] table[row sep=crcr]{%
x	y\\
5	11.027221341301\\
};
\addplot[only marks,line width=0.9pt,  mark=square, mark options={}, mark size=3.7pt, draw=black, forget plot] table[row sep=crcr]{%
x	y\\
6	9.12421719889121\\
};
\addplot[only marks,line width=0.9pt,  mark=square, mark options={}, mark size=3.7pt, draw=black, forget plot] table[row sep=crcr]{%
x	y\\
8	15.5751841785941\\
};
\addplot[only marks,line width=0.9pt,  mark=square, mark options={}, mark size=3.7pt, draw=black, forget plot] table[row sep=crcr]{%
x	y\\
10	23.7185996932192\\
};
\addplot[only marks,line width=0.9pt,  mark=square, mark options={}, mark size=3.7pt, draw=black, forget plot] table[row sep=crcr]{%
x	y\\
0	7.82\\
};
\addplot[only marks,line width=0.9pt,  mark=square, mark options={}, mark size=3.7pt, draw=black, forget plot] table[row sep=crcr]{%
x	y\\
2	8.68\\
};
\addplot[only marks,line width=0.9pt,  mark=square, mark options={}, mark size=3.7pt, draw=black, forget plot] table[row sep=crcr]{%
x	y\\
4	10.84\\
};
\addplot[only marks,line width=0.9pt,  mark=square, mark options={}, mark size=3.7pt, draw=black, forget plot] table[row sep=crcr]{%
x	y\\
6	14.55\\
};
\addplot[only marks,line width=0.9pt,  mark=square, mark options={}, mark size=3.7pt, draw=black, forget plot] table[row sep=crcr]{%
x	y\\
7	15.94\\
};
\addplot[only marks,line width=0.9pt,  mark=square, mark options={}, mark size=3.7pt, draw=black, forget plot] table[row sep=crcr]{%
x	y\\
9	22.54\\
};
\addplot[only marks,line width=0.9pt,  mark=square, mark options={}, mark size=3.7pt, draw=black, forget plot] table[row sep=crcr]{%
x	y\\
8	21.65\\
};
\addplot [color=mycolor4,line width=0.9pt,  forget plot]
  table[row sep=crcr]{%
0	4.73205627280729\\
11	4.73205627280729\\
};
\addplot [color=mycolor4,line width=0.9pt,  dashdotted, forget plot]
  table[row sep=crcr]{%
0	7.63473859029459\\
0.5	7.63473859029459\\
};
\addplot [color=mycolor4,line width=0.9pt,  dashdotted, forget plot]
  table[row sep=crcr]{%
0.5	7.87182913433877\\
1.5	7.87182913433877\\
};
\addplot [color=mycolor4,line width=0.9pt,  dashdotted, forget plot]
  table[row sep=crcr]{%
1.5	8.36424453137129\\
2.5	8.36424453137129\\
};
\addplot [color=mycolor4,line width=0.9pt,  dashdotted, forget plot]
  table[row sep=crcr]{%
2.5	9.13974705216197\\
3.5	9.13974705216197\\
};
\addplot [color=mycolor4,line width=0.9pt,  dashdotted, forget plot]
  table[row sep=crcr]{%
3.5	10.2266547365514\\
4.5	10.2266547365514\\
};
\addplot [color=mycolor4,line width=0.9pt,  dashdotted, forget plot]
  table[row sep=crcr]{%
4.5	11.6506207539372\\
5.5	11.6506207539372\\
};
\addplot [color=mycolor4,line width=0.9pt,  dashdotted, forget plot]
  table[row sep=crcr]{%
5.5	13.4358501223517\\
6.5	13.4358501223517\\
};
\addplot [color=mycolor4,line width=0.9pt,  dashdotted, forget plot]
  table[row sep=crcr]{%
6.5	15.6076495582769\\
7.5	15.6076495582769\\
};
\addplot [color=mycolor4,line width=0.9pt,  dashdotted, forget plot]
  table[row sep=crcr]{%
7.5	18.1950615051224\\
8.5	18.1950615051224\\
};
\addplot [color=mycolor4,line width=0.9pt,  dashdotted, forget plot]
  table[row sep=crcr]{%
8.5	21.2330052735993\\
9.5	21.2330052735993\\
};
\addplot [color=mycolor4,line width=0.9pt,  dashdotted, forget plot]
  table[row sep=crcr]{%
9.5	24.7622656455279\\
10.5	24.7622656455279\\
};
\end{axis}
\end{tikzpicture}%

%% file: widths.tex
%
%
\definecolor{mycolor1}{rgb}{0.00000,0.44700,0.74100}%
\definecolor{mycolor2}{rgb}{0.85000,0.32500,0.09800}%
\definecolor{mycolor3}{rgb}{0.63500,0.07800,0.18400}%
\begin{tikzpicture}

\begin{axis}[%
  width=4.2in,
  height=3in,
at={(0.758in,0.481in)},
scale only axis,
xmin=0,
xmax=10,
tick align=outside,
xlabel style={font=\color{white!15!black}},
xlabel={$l$},
ymode=log,
ymin=0.001,
ymax=10000,
yminorticks=true,
ylabel style={font=\color{white!15!black}},
ylabel={$\Gamma$ [keV]},
axis background/.style={fill=white}
]
\addplot [color=mycolor1,line width=0.9pt, mark=o, mark options={solid, mycolor1}, forget plot]
  table[row sep=crcr]{%
1	0.0015\\
3	16\\
5	320\\
7	1000\\
9	2200\\
};
\addplot [color=mycolor1,line width=0.9pt, dashed, mark=o, mark options={solid, mycolor1}, forget plot]
  table[row sep=crcr]{%
1	0.028\\
3	8.2\\
5	150\\
7	110\\
9	500\\
};
\addplot [color=mycolor2,line width=0.9pt, mark=o, mark options={solid, mycolor2}, forget plot]
  table[row sep=crcr]{%
0	850\\
2	1200\\
4	2300\\
6	4500\\
8	7100\\
10	8900\\
};
\addplot [color=mycolor2,line width=0.9pt, dashed, mark=o, mark options={solid, mycolor2}, forget plot]
  table[row sep=crcr]{%
0	800\\
2	800\\
4	350\\
};
\addplot [color=mycolor3,line width=0.9pt, mark=o, mark options={solid, mycolor3}, forget plot]
  table[row sep=crcr]{%
6	0.19\\
8	38\\
10	330\\
};
\addplot [color=mycolor3,line width=0.9pt, dashed, mark=o, mark options={solid, mycolor3}, forget plot]
  table[row sep=crcr]{%
6	0.11\\
8	100\\
};
\end{axis}
\end{tikzpicture}%